\documentclass[aps,prd,amsmath,amssymb,preprintnumbers,onecolumn,11pt,nofootinbib]{revtex4}
\usepackage[a4paper,left=20mm,right=20mm,top=25mm,bottom=25mm]{geometry}
\usepackage{amsmath,amssymb}
\usepackage{mathrsfs}
\usepackage{graphicx}
\usepackage{color}
\usepackage{subfigure}
\usepackage{fancyhdr}
\usepackage{multirow}
\usepackage{float}
\usepackage{epsfig}
\usepackage{amsfonts}
\usepackage{bm}
\usepackage{xcolor}

\definecolor{CP3}{cmyk}{0,0.88,0.77,0.40}

\newcommand{\La}{\Lambda}

\newcommand{\be}{\begin{equation}}
\newcommand{\ee}{\end{equation}}
\newcommand{\ba}{\begin{eqnarray}}
\newcommand{\ea}{\end{eqnarray}}
\newcommand{\tg}{\tilde{g}}

\renewcommand{\(}{\left(}
\renewcommand{\)}{\right)}

\begin{document}

\title{Generalized Conformal Transformation and Inflationary Attractors}

\author{Khamphee Karwan$^{1,2}$}
\email{khampheek@nu.ac.th} 
\author{Phongpichit Channuie$^{3,4,5}$}
\email{channuie@gmail.com}
\affiliation{\footnotesize $^{1}${The Institute for Fundamental Study \lq\lq The Tah Poe Academia Institute\rq\rq, \\Naresuan University, Phitsanulok 65000, Thailand}}
\affiliation{\footnotesize $^{2}${Thailand Center of Excellence in Physics, Ministry of Education, Bangkok 10400, Thailand}}
\affiliation{\footnotesize $^{3}${College of Graduate Studies, Walailak University, Nakhon Si Thammarat, 80160 Thailand}}
\affiliation{\footnotesize $^{4}${School of Science, Walailak University, Thasala, Nakhon Si Thammarat, 80160, Thailand}}
\affiliation{\footnotesize $^{5}${Bioscience \& Artificial Intelligence Unit, Walailak University, Nakhon Si Thammarat, 80160, Thailand}}

\begin{abstract}
We investigate the inflationary attractors in models of inflation  inspired from general conformal transformation of general scalar-tensor theories to the Einstein frame. The coefficient of the conformal transformation in our study depends on both the scalar field and its kinetic term. Therefore  the relevant scalar-tensor theories display the subset of the class I of the degenerate higher order scalar-tensor theories in which both the scalar field and its kinetic term can non-minimally couple to gravity. We find that if the conformal coefficient $\Omega$ takes a multiplicative form such that $\Omega \equiv w(\phi)W(X)$ where $X$ is the kinetic term of the field $\phi$, the theoretical predictions of the proposed model can have usual universal attractor independent of any functions of $W(X)$. For the case where $\Omega$ takes an additive form, such that $\Omega \equiv w(\phi) + k(\phi) \Xi(X)$, we find that there are new $\xi$ attractors in addition to the universal ones. We analyze the inflationary observables of these models and compare them to the latest constraints from the Planck collaboration. We find that the observable quantities associated to these new $\xi$ attractors do not satisfy the  constraints from Planck data at strong coupling limit.
 \\[4mm]
\end{abstract}

\maketitle

\section{Introduction}

The mechanism of cosmic inflation is a conceivable framework when one wants to describe the universe at very early times.  It can nicely address a number of issues of stadard Big
Bang cosmology. More concretely, it paves the treatment of primordial fluctuations resulting
in the large scale structures and the anisotropy in the temperature of the cosmic microwave
background (CMB) observed today. In the simplest version of the models, we require the presence of a scalar degree
of freedom (inflaton), either as a fundamental scalar field, e.g. a Higgs field \cite{Bezrukov:2007ep,Barvinsky:2008ia,Bezrukov:2008ut,Bezrukov:2008ej,Bezrukov:2009db,Barbon:2009ya} or a composite field \cite{Evans:2010tf,Channuie:2011rq,Bezrukov:2011mv,Channuie:2012bv,Channuie:2016iyy,Samart:2018ucu} (or even incorporated into gravity itself), in
general as an effective scalar degree of freedom. More recently, a broad class of inflationary models, dubbed cosmological
attractors \cite{Kallosh:2013hoa,Kallosh:20131,Kallosh:20132,Kallosh:2014rga} has attracted a lot of attention. Cosmological attractor scenarios for the inflationary models have been developed in the past few years \cite{Galante:2014ifa,Cecotti:2014ipa,Yi:2016jqr,Carrasco:2015rva,Carrasco:2015uma}.

Interestingly, the cosmological $\alpha$-attractors constitute most of the existing inflationary models with
plateau-like potentials. These include the Starobinsky model and some generalized versions of the Higgs inflation. Regarding the $\alpha$-attractors, the flatness of the inflaton potential is achieved and protected
by the existence of a pole in the kinetic term of the scalar field. Moreover, at large-field values, any
non-singular inflaton potential acquires a universal plateau-like form when performing the (conformal)
transformation. Regarding the hyperbolic geometry and the flatness of the Kahler potential in the supergravity
context, the universal behaviors of these theories make very similar cosmological predictions
preserving good agreement with the current observational data \cite{Ade:2015lrj}. This class of models has certain universal predictions for the important cosmological observables, i.e. scalar spectral index ($n_{s}$) and tensor-to-scalar ratio ($r$). It has been shown that the non-minimal coupling between inflaton and gravity in the strong coupling limit can lead to attractor which the observational quantities are the same as  the universal $\alpha$ attractors \cite{Kallosh:20131,Kallosh:20132}.
The general consideration for the relations between the inflationary attractor due to the non-minimal coupling, namely $\xi$ attractors, and the $\alpha$ attractors is presented in \cite{Galante:2014ifa}.

In the present work, we extend analysis in the existing literature by considering the cases where the non-minimal coupling is also in the form of non-minimal kinetic coupling such that the term $k(\phi) f(X) R$ appears in the action.
Here, $k(\phi)$ is an arbitrary function of the inflaton $\phi$, $f(X)$ is an arbitrary function of $X$, $X = - \partial_\mu\phi \partial^\mu\phi /2$ is the kinetic term of the inflaton field and $R$ is the Ricci scalar.
In general such non-minimal coupling arises by applying the general conformal transformation, in which the conformal coefficient depends on both the scalar field and its kinetic term, to the Einstein-Hilbert action. In Sec.(\ref{sec1}), we construct an  action in the Einstein frame that is conformally equivalent to scalar-tensor theories with a general non-minimal coupling  using the general conformal transformation.
We investigate inflationary attractors in the presence of the general non-minimal coupling using the action in the Einstein frame based on the assumption that observable quantities are frame-invariant.
The two-case scenarios are considered. In Sec.(\ref{multi1}), we concentrate on the multiplicative form of the generalized conformal factor, i.e. $\Omega=\Omega(X,\phi)=w(\phi)W(X)$. We show whether the attractors those found in literature can exist in our models, and then review some essential ideas of the inflationary attractors as well as calculations of cosmological observables, i.e. $n_{s}$, and $r$ considering both hyperbolic tangent potential and exponential potential. In Sec.(\ref{add}), we choose  the additive form of the generalized conformal factor, i.e. $\Omega=\Omega(X,\phi)=w(\phi)+k(\phi)\Xi(X)$ which in some situations can be viewed as generalization of the multiplicative form models.
We compute the cosmological observables for hyperbolic tangent potential, and consider theoretical predictions in the weak and strong coupling limits which are equivalent to large and small $\alpha$ limits in our setup, respectively. In Sec.(\ref{planck}), we compare the obtained results of the  cosmological observables with recent Planck
2015 data. Finally, we present our conclusion in the last section.

\section{General conformal transformation and action in the Einstein frame}
\label{sec1}
Let us first consider a general conformal transformation in which the relation between a new metric, ${\tilde g}_{\mu\nu}$, and the old one, $g_{\mu\nu}$,  takes the form:
\be
\label{trans}
 \tg_{\mu\nu} = \Omega(X,\phi) g_{\mu\nu}\,.
\ee
According to this transformation, the determinant between the two metrices yields
\be
{\cal J}_g\equiv \frac{\sqrt{-\tg}}{\sqrt{-g}}=\Omega^2\,,
\ee
and a relation between kinetic terms in different frames is
\be
\tilde X \equiv - \frac 12 \tg^{\mu\nu}\partial_\mu\phi \partial_\nu \phi = \frac{X}{\Omega}\,.
\label{xbarx}
\ee
Applying the transformation in Eq.~(\ref{trans}) to the Einstein-Hilbert action,
\be
S_E = \int d^4x \sqrt{-\tg} \frac 12 \tilde{R}\,,
\label{se}
\ee
we get \cite{Zumalaca}
\be
S_J = \int d^4x \sqrt{-g}\left [\frac{1}{2} \Omega R + \frac 34 \frac{\Omega_\phi^2}{\Omega} \partial_\alpha\phi\partial^\alpha\phi
+ \frac 3{2\Omega} \Omega_\phi\Omega_X \partial_\alpha\phi\partial^\alpha X
+ \frac 34 \frac{\Omega_X^2}{\Omega} \partial_\alpha X \partial^\alpha X\right ]\,,
\label{sj}
\ee
where subscripts ${}_\phi$ and ${}_X$ denote derivative with respect to $\phi$ and $X$, respectively. Here, we set the reduced Planck mass $M_{P}=(8\pi G)^{-1/2}=1$.
We now add kinetic term $- \sqrt{- \tg} h(\phi,X) \tg^{\alpha\beta} \partial_\alpha \phi \partial_\beta \phi / 2$ to the Einstein-frame action in Eq.~(\ref{se}).
Under the transformation given in Eq.~(\ref{trans}),
this kinetic term gives $- \sqrt{-g}\Omega h g^{\alpha\beta}\partial_\alpha\phi\partial_\beta\phi/2$ in the Jordan-frame action.
Let us define the kinetic term of scalar field in the Jordan Frame as
$- \sqrt{-g}G(\phi,X) g^{\alpha\beta}\partial_\alpha\phi\partial_\beta\phi / 2$.
Hence, we have
\be
- G(\phi,X) = \frac 32 \frac{\Omega_\phi^2}{\Omega} - \Omega h(\phi,X)\,,
\ee
and therefore
\be
h = \frac{G + 3 \Omega_\phi^2 / (2 \Omega)}{\Omega}\,.
\label{coefkin}
\ee
Based on the above analysis,
we conclude that under the transformation given in Eq.~(\ref{trans}) the action in the Einstein frame
\be
S_E = \int d^4x \sqrt{-\tg}\left[ \frac 12 \tilde{R}
- \frac{G\Omega + 3 \Omega_\phi^2/ 2}{2 \Omega^2}\tg^{\alpha\beta}\partial_\alpha\phi\partial_\beta\phi
\right]\,,
\label{semod}
\ee
becomes
\be
S_J = \int d^4x \sqrt{-g}\left[\frac{1}{2} \Omega R - \frac 12 G(\phi,X) \partial_\mu\phi\partial^\mu\phi
+ \frac 3{2\Omega} \Omega_\phi\Omega_X \partial_\alpha\phi\partial^\alpha X
+ \frac 34 \frac{\Omega_X^2}{\Omega} \partial_\alpha X \partial^\alpha X\right]\,.
\label{sjmod}
\ee
The potential term for the scalar field in the Einstein frame can be obtained by adding the term $- \Omega^2 V_E(\phi)$ in the Jordan-frame action. Thus under the general conformal transformation, the action in the Jordan frame
\be
S_J = \int d^4x \sqrt{-g}\left[\frac{1}{2} \Omega R + G(\phi,X) X
- \Omega^2 V_E(\phi)
+ \frac 3{2 \Omega} \Omega_\phi\Omega_X \partial_\alpha\phi\partial^\alpha X
+ \frac 34 \frac{\Omega_X^2}{\Omega} \partial_\alpha X \partial^\alpha X\right]\,,
\label{sjmodv}
\ee
is equivalent to the Einstein-frame action
 \be
S_E = \int d^4x \sqrt{-\tg}\left[ \frac 12 \tilde{R}
+ \frac{G\Omega + 3 \Omega_\phi^2 / 2}{\Omega^2}\tilde X
- V_E(\phi)
\right]\,.
\label{semodv}
\ee
We note that the coefficients $G(\phi,X)$ and $\Omega(\phi,X)$ in the above Einstein-frame action depend on kinetic terms $X$ in general.
We will consider in the subsequent sections the cases where the $X$-dependent terms in the Einstein-frame action can cancel each other or can be transformed to $\tilde X$.

The combination of the second and third terms in the action (\ref{sjmodv}) is the Lagrangian of K-inflation,
which can be defined as $L_2 \equiv G X - \Omega^2 V_E(\phi)$.
Using the definition $3 \Omega_\phi\Omega_X / (2 \Omega)\equiv F(\phi,X) + F_X(\phi,X) X$,
the third term in the action can be integrated by parts yielding the cubic galileon term $- F \Box\phi$.
The fourth term in the action is a subset of the degenerate higher order scalar-tensor theories (DHOST),
 so that it does not lead to Ostrogradski instability \cite{Langlois:2015cwa, Langlois:2015skt, Crisostomi:2016tcp, Crisostomi:2016czh, Achour:2016rkg, Langlois:2017}.
Due to the existence of this term, the theory  described by the action (\ref{sjmodv})  belongs to  the class I of DHOST theory in which 
the Laplacian instabilities emerging from negative sound speed of the cosmological perturbations  disappear \cite{Langlois:2017}.
Moreover, this theory satisfies the conditions for which propagation speed of gravitational waves equals to speed of light \cite{Langlois:2017, Crisostomi:2017}.

In principle, physical quantities predicted from inflationary model described by action (\ref{sjmodv})  are the same as those obtained from the action in Eq.~(\ref{semodv}).
However, to explicitly verify this statement, the predictions such as spectral indices and tensor-to-scalar ratio of the perturbation amplitudes from DHOST theories have to be studied in which we leave for future investigation.
Although the theoretical predictions are  expected to be frame invariant, the comparisons between predictions from inflationary models and results from observational data require relation
 between the predicted quantities and the number of e-folding of inflation which is frame dependent \cite{Tsujikawa:2004my, Karam:2017zno}.
To investigate how the number of e-folding depends on the frame,
we suppose that the background metric is spatially flat  Friedmann-Lema\^{i}tre-Robertson-Walker (FLRW) metric given by
\be
ds^2 = - dt^2 + a^2(t) \delta_{ij} dx^i dx^j\,,
\label{flrw}
\ee
where $a$ is the cosmic scale factor and $\delta_{ij}$ is the Kronecker delta. Therefore Eq.~(\ref{trans}) yields
\be
\tilde{a} = \sqrt{\Omega(X,\phi)} a\,,
\ee
where $\tilde{a}$ and $a$ are the cosmic scale factors in the Einstein and Jordan frames, respectively.
Hence, the relation between the number of e-folding for different frames is given by
\be
\tilde{N} = N  + \frac 12 \ln\(\frac{\Omega_{\rm end}}{\Omega_N}\)\,,
\label{tn2n}
\ee
where subscript ${}_{\rm end}$ denotes evaluation at the end of inflation,
while subscript ${}_N$ represents the quantities evaluated at the horizon crossing at e-folding $N$ of the observed CMB modes.
Since the function $\Omega$ describes non-minimal coupling in the Jordan frame,
it can be generally written in the form
\be
\Omega(X,\phi) = 1 + \xi F(\phi, X)\,,
\label{omega-gen}
\ee
where $\xi$ is the non-minimal coupling constant.
Inserting Eq.~(\ref{omega-gen}) into Eq.~(\ref{tn2n}),
we get
\be
\tilde{N} = N  + \frac 12 \ln\(\frac{1 + \xi F_{\rm end}}{1 + \xi F_N}\)\,
= N + \left\{\begin{array}{cc}
\frac 12 \xi \(F_{\rm end} - F_N\) & \quad\mbox{for}\quad \xi \ll 1
\\
\frac 12 \ln\(\frac{F_{\rm end}}{F_N}\) & \quad\mbox{for}\quad \xi \gg 1
\end{array}\right.\,.
\label{tn2n2}
\ee
It clearly follows from the above equation that if $\phi$ is supposed to  slowly evolve during inflation,
we can make an approximation $\tilde{N} \simeq N$ when $\xi$ is sufficiently small. In the case of large $\xi$ limit, an approximation $\tilde{N} \simeq N$ is valid when the function $F$ changes slowly during inflation.
To estimate how much the function $F$ changes during inflation,
we compute evolution equations for the background Universe. Starting from the action given in Eq.~(\ref{sjmodv}), we insert the metric in Eq.~(\ref{flrw}) into the action and vary the action with respect to components of the metric. Here varying the action with respect to $a$, we obtain
\ba
0 &=& 2 \dot H \Omega + 3 H^2 \Omega + \ddot\phi \Omega_\phi + \ddot\phi^2 \Omega_X + \dddot\phi \dot\phi \Omega_X 
+ 2 \dot\phi H \(\Omega_\phi + \ddot\phi \Omega_X\) - \Omega^2 V_E \nonumber\\&&+ G X + 2 \Omega_{\phi\phi} X + 4 \ddot\phi \Omega_{\phi X} X 
- \frac 3{\Omega} \Omega_\phi \Omega_X X \ddot\phi - \frac 3{2 \Omega} \Omega_X^2 X \ddot\phi^2 + 2 \ddot\phi^2 \Omega_{XX} X\,,
\label{dh}
\ea
where $H \equiv \dot a / a$ is the Huble parameter and a dot denotes derivative with respect to time $t$.
To vary the action with respect to the (00) component of the metric,
we introduce an auxiliary function $\eta(t)$ in which we find the replacement of $- dt^2$ in Eq.~(\ref{flrw}) with $- \eta^2(t) dt^2$. Vary the action with respect to $\eta$, setting $\eta = 1$ in the obtained result,
and then eliminating $\ddot a$ from the resulting evolution equation by Eq.~(\ref{dh}),
we get
\ba
0 &=& -2 \Omega^4 V_E + 2 \Omega^3 \Omega_X V_E X 
- 3 \Omega_X \(-4 \Omega_\phi^2 + 2 \ddot\phi \Omega_\phi \Omega_X + \ddot\phi^2 \Omega_X^2\) X^2 
- 2 \Omega^2 X \(G + 2 G_X X\) \nonumber\\&&+ 6 H^2 \Omega^2 \(\Omega - \Omega_X X\) 
+ 6 \dot\phi H \Omega \(\Omega_\phi + \ddot\phi \Omega_X\) \(\Omega - \Omega_X X\) \nonumber\\&&+ 
  3 \Omega X \(2 \ddot\phi \Omega_\phi \Omega_X + \ddot\phi^2 \Omega_X^2 - 4 \Omega_\phi \Omega_{\phi X} X + 2 G \Omega_X X\)\,, 
\label{h2}
\ea
Combining Eq.~(\ref{dh}) with Eq.~(\ref{h2}),
we can write the expression for $ - \dot H / H^2$ which is the slow-roll parameter in terms of dimensionless parameters as
\ba
- \frac{\dot H}{H^2} &=& 
 x_6 \(2 x_4 x_6 - 3 x_2^2 x_6 - x_2 \(1 - x_6 - x_7\)\) 
-\frac 12 \(x_1 - x_5 - x_1 x_6 + 6 x_1 x_2 x_6 - 4 x_3 x_6\)
\nonumber\\
&&- \frac 1{1 - x_2} \left[\frac32 x_1 \(x_1 x_2 - x_3\) 
- \(1 - 2 x_2\) x_{g0} - x_{g0} x_{g2}
\right]\,,
\label{dh2h2}
\ea
where
\ba
x_1 &\equiv & \frac{\Omega_\phi \dot\phi}{H \Omega}\,,\quad
x_2 \equiv \frac{\Omega_X X}{\Omega}\,,\quad
x_3 \equiv \frac{\Omega_{\phi X} \dot\phi X}{H \Omega}\,,\quad
x_4 \equiv \frac{\Omega_{XX}X^2}{\Omega}\,,\quad
x_5 \equiv \frac{\Omega_{\phi\phi} \dot\phi^2}{H^2 \Omega}\,,
\nonumber\\
x_6 &\equiv & \frac{\ddot\phi}{\dot\phi H}\,,\quad
x_7 \equiv \frac{\dddot\phi}{\ddot\phi H}\,,\quad
x_{g0} \equiv \frac{G X}{H^2 \Omega}\,,\quad
x_{g2} \equiv \frac{G_X X^2}{H^2 \Omega}\,.
 \ea
At the leading order,
Eq.~(\ref{dh2h2}) gives
\ba
- \frac{\dot H}{H^2}  &\simeq &
- x_2 x_6 -\frac 12 x_1 
+ \frac 12 x_5 + \frac{1 - 2 x_2}{1 - x_2} x_{g0}
\nonumber\\
&=&
- \frac{\dot{\Omega}}{2 H \Omega} 
+ \( \frac{\Omega_{\phi\phi}}{\Omega} + \frac{1 - 2 x_2}{1 - x_2} \frac{G}{\Omega}\) \frac{X}{H^2}\,. \label{21}
\ea
The above equation suggests that $- \dot\Omega / (2 H \Omega)$ as well as the remaining term on the right-hand-side of (\ref{21}) should be in the same order as $ - \dot H / H^2$, i.e., $\dot\Omega / (2 \Omega) \lesssim \dot H / H$.
Hence, for a large $\xi$ limit,
we have $\dot F / (2 F) \lesssim\dot H / H$,
implying $F \sim H^{2 s} $ where $s \lesssim 1$.
Inserting this result into Eq.~(\ref{tn2n2}),
we get
\be
\tilde{N} \simeq N + s \ln\(\frac{H_{\rm end}}{H_N}\)\,.
\label{ndif}
\ee
From the PLANCK results \cite{Akrami:2018odb},
we have $H_N \lesssim 2.7 \times 10^{-5} M_p$. Suppose that the Hubble parameter during inflation is almost constant. Hence we can approximately ignor the second term on the RHS of Eq.~(\ref{ndif}),
and therefore we have $\tilde{N} \simeq N$. Hence, the predicted quantities in terms of number of e-folding from the action in Eqs.~(\ref{sjmodv}) and (\ref{semodv}) are approximately the same in the strong ($\xi \gg 1$) and weak ($\xi \ll 1$) limits.

In the following consideration, we will investigate the attractor of the theoretical predictions from the inflationary model described by the action (\ref{semodv}).
Based on the discussion in the preceding paragraph, the inflationary attractors in the models described by the action (\ref{semodv})
should imply the same attractors appearing  in the subclass of DHOST theories described by the action (\ref{sjmodv}) in the strong and weak coupling limits. These attractors are consequences of general non-minimal coupling associated with general conformal transformation which are the main interests of this work.
Actually non-minimal coupling can also be associated with another type of frame transformation called disformal transformation \cite{Zumalaca}.
Some of subclasses of DHOST theories can be transformed to the Einstein frame using the disformal transformation.
The kinetic terms of scalar field in the resulting action in the Einstein frame should also take non-canonical form.
Hence, in this section we consider the general conformal transformation between the Jordan and Einstein frames to ensure that the Einstein action using in our calculation represents effects of non-minimal coupling associated with the general conformal transformation.

\section{Multiplicative form}
\label{multi1}
We first consider the case where $\Omega$ has an multiplicative form, such that
\be
\Omega(\phi, X) = w(\phi)\,W(X)\,.
\label{wmulti}
\ee
To make our consideration independent of the form of $W(X)$,
we set $G(\phi, X) = g(\phi)W(X)$ and then Eq.~(\ref{semodv}) becomes
 \be
S_E = \int d^4x \sqrt{-\tg}\left[ \frac 12 \tilde{R}
- \frac{gw + 3 w_\phi^2 / 2}{2 w^2}\tg^{\alpha\beta}\partial_\alpha\phi\partial_\beta\phi
- V_{E}(\phi)
\right]\,.
\label{semodv-mt}
\ee
For suitable choices of field-redefinition,
inflationary models described by the above action should have usual inflationary attractor as those found in the literature. In terms of the canonical normalized field $\psi$, the above action takes the form
 \be
S_E = \int d^4x \sqrt{-\tg}\left[ \frac 12 \tilde{R}
- \frac 12 \tg^{\alpha\beta}\partial_\alpha\psi\partial_\beta\psi
- V(\psi)
\right]\,,
\label{ca}
\ee
where
 \be
d\psi^{2} = \left(\frac{gw + 3 w_\phi^2 / 2}{w^2}\right)d\phi^{2}\,.
\label{dpsdph}
\ee
Since the action (\ref{semodv-mt}) is similar to the action in the Einstein frame for scalar-tensor theories with non-minimal coupling term $w(\phi)$,
we set $w(\phi) \equiv 1 + \xi f(\phi)$ with a dimensionless coupling constant $\xi$ and an arbitrary function $f(\phi)$.
To obtain exact relation between $\psi$ and $w(\phi)$,
the relation between $w(\phi)$ and the kinetic coupling $g(\phi)$ is supposed to satisfy the following condition  \cite{Galante:2014ifa}, 
 \be
g(\phi) = \frac{1}{4\xi}\left(\frac{ w_\phi^2 }{w}\right)\,,
\label{tc1a}
\ee
then Eq.~(\ref{dpsdph}) gives $\psi=\sqrt{3\alpha/2}\ln w(\phi)$. This yields 
 \be
w(\phi) = \exp(\sqrt{2/3\alpha}\psi)\,,
\label{tca}
\ee
where $\alpha=1+(6\xi)^{-1}$.
Based on the above exact relation between $\psi$ and $w(\phi)$,
the action (\ref{ca}) will be independent from $w(\phi)$ if $V_E(\phi)$ is a function of $w(\phi)$.
The slow roll parameters, $\epsilon,\eta$, and the number of e-folding, $N$, have the same forms as the standard slow-roll paradigm, and they read
\begin{eqnarray}
\epsilon = \frac{1}{2} \left( \frac{1}{V}\frac{dV}{d \psi} \right)^2,\,
\eta = \left( \frac{1}{V}\frac{d^2V}{d \psi^2} \right),\,
N = \int _{\psi_{\rm end}} ^{\psi_N} d \psi  \frac{V}{dV/d\psi}\,,
\label{slow-roll}
\end{eqnarray}
where $\psi_{\rm end}$ is the value of $\psi$ at the end of inflation,
and $\psi_N$ is the value of $\psi$ at given $N$.

We can test our predictions with the experimental results by using the relative strength of the tensor perturbation, i.e. the tensor-to-scalar ratio $r$ and the spectral index of curvature perturbation $n_{s}$. In terms of the slow-roll
parameters, these observables are written as
\begin{eqnarray}
r=16\,\epsilon_{N}\,,
\quad 
n_{s}=1-6\,\epsilon_{N}+2\eta_{N}\,.
\label{rns-st}
\end{eqnarray}
Regarding the relation in Eq.(\ref{tca}), we consider
\ba
V_E(\phi)=V_{0} \left[\frac{w(\phi)-1}{w(\phi)+1}\right]^{n}\,,
\label{mul2}
\ea
which leads to
\be
V(\psi) = V_{0}\tanh^{n}\Big(\frac{\psi}{\sqrt{6\alpha}}\Big)\,,
\label{tE1}
\ee
which is a well-known attractor potential and note explicitly that Ref.\cite{Kallosh:2013yoa} gives $r$ and $n_s$ shown below. Since the potential takes the form of hyperbolic tangent,
this class of models is called T-model \cite{Ferrara:2013rsa,Kallosh:2013yoa}.
Having used the effective potential in Eq.~(\ref{tE1}), the observable quantities given in Eq.~(\ref{rns-st}) can be written in terms of $N$ as \cite{Kallosh:2013yoa}
\begin{eqnarray}
r &=&\frac{12 n \alpha}{nN^{2}+{\cal G}(\alpha) N+3n \alpha/4}\,,
\label{r-tm}\\
n_{s}&=&\frac{n (4 (-2+N) N-3 \alpha )+4(-1+N) \sqrt{3\alpha  \left(n^2+3 \alpha \right)}}{4 n N^2+3 n \alpha +4N \sqrt{3\alpha  \left(n^2+3 \alpha \right)}}\\
&=&\frac{1-\frac{2}{N}-\frac{3\alpha}{4N^{2}}+\frac{1}{2N}\Big(1-\frac{1}{N}\Big){\cal G}(\alpha)}
{1+\frac{1}{2N}{\cal G}(\alpha)+\frac{3\alpha}{4N^{2}}}\,,
\label{ns-tm}
\end{eqnarray}
where ${\cal G}(\alpha)=\sqrt{3\alpha}\sqrt{(3\alpha+n^{2})}$.
To the lowest order in the slow-roll approximation, the inflationary predictions in terms of the
number of e-foldings in the Einstein frame parameters
for this model read:
\begin{eqnarray}
n_{s} &=& 1-\frac{2n + 4}{4 N+n}, 
\quad\,\,\,
r=\frac{16 n}{4 N+n}
 \text{for } \alpha \gg 1 \& \alpha \gg n, 
\label{attract1}\\
    n_{s} &=& 1-\frac{2}{N},  \quad\quad\quad
r= \frac{12 \alpha }{N^2} 
 \text{for } \alpha \ll 1\,.
\label{uniattrat}
\end{eqnarray}
The above expressions for $n_s$ and $r$ in the large and small $\alpha$ limits are computed by treating $\alpha$ as a free parameter which 
controls the slope of $V(\psi)$.
From the definition of $\alpha$ in terms of the coupling constant $\xi$,
we take $\alpha \to \infty$ in the weak coupling $\xi \ll 1$ limit and $\alpha \to 1$ in the strong $\xi \gg 1$ limit. We will see in the numerical investigation displaying in Fig.(\ref{figlrns}) that in the strong coupling limit ($\alpha = 1$), the observable quantities converge to the universal attractor regime in Eq.~(\ref{uniattrat}) \cite{Kallosh:2013yoa,Ferrara:2013rsa, Galante:2014ifa}. This regime corresponds to the part
of the $n_{s}- r$ plane favored by the Planck data \cite{Ade:2013zuv}. For small coupling limit, the predictions  converge to  Eq.~(\ref{attract1}) if $n$ is replaced by $2n$. Moreover, regarding the relation in Eq.(\ref{tca}), the potential of the field $\psi$ takes the exponential form, namely E-model \cite{Ferrara:2013rsa,Kallosh:2013yoa},
if we set $V_{E}(\phi)=V_{0}\left[1-w^{-1}(\phi)\right]^{n}$.
This form of $V_{E}(\phi)$ yields
\be
V(\psi) = V_{0}\left[1-\exp(-\sqrt{2/3\alpha}\psi)\right]^{n}\,. 
\label{E1}
\ee
For this form of the potential,
it is difficult to write the time-varying parts of the inflationary predictions $r$ and $n_s$ solely in terms of the number of e-folding as in Eqs.~(\ref{r-tm}) and (\ref{ns-tm}).
Hence, we consider the inflationary predictions for this case in the large and small $\alpha$ limits.
In the large $\alpha$ limit, the above potential coincides with the simplest chaotic inflation model with $\psi^n$-potential. In the limit $\alpha \gg 1$, i.e.$\sqrt{2/3\alpha}\ll 1$, we have
\be
V(\psi) = V_{0}\left[1-\exp(-\sqrt{2/3\alpha}\psi)\right]^{n}=V_{0}\left[1-\exp(-\sqrt{2/3\alpha}\psi)\right]^{n}\simeq \frac{2}{3\alpha}V_{0}\psi^{n}\equiv {\tilde V}_{0}\psi^{n}. 
\label{El1}
\ee
For this potential, the slow-roll parameters take the form
\begin{eqnarray}
\epsilon = \frac{n^{2}}{2\psi ^2} ,\,\,\,\eta = \frac{n(n-1)}{\psi ^2} .
\end{eqnarray}
Slow-roll inflation terminates when $\epsilon = 1 $, so the field value at the end of inflation reads
\begin{eqnarray}
\epsilon = \frac{n^{2}}{2\psi ^2}=1 \longrightarrow \psi_{\rm end} = \frac{n}{\sqrt{2}} .
\end{eqnarray}
The number of e-folding for the change of the field $\psi$ from $\psi_{N}$ to $\psi_{\rm end}$ is given by
\begin{eqnarray}
N+N_{\rm e} = \frac{\psi_N ^2}{2n}\quad {\rm with}\quad N_{\rm e} =\frac{\psi ^2_{\rm end}}{2n} .
\end{eqnarray}
Therefore, in terms of $N$, the values of $n_{s}$ and $r$ for the large $\alpha$ limit are given by
\begin{eqnarray}
n_{s}=1-\frac{2 n + 4}{4 N + n}\,,\quad
r = \frac{16n}{4N+n}\,, 
\quad\text{for }\quad \alpha \gg 1.
\label{attract1-e}
\end{eqnarray}
However, in the small $\alpha$ limit, i.e. $\alpha \ll1$,
the potential in Eq.~(\ref{E1}) becomes
\be
V(\psi) \simeq V_{0}\left[1- n \exp(-\sqrt{2/3\alpha}\psi)\right]\,.
\label{E1-small}
\ee
For this potential, the slow-roll parameters are
\begin{eqnarray}
\epsilon \simeq \frac{n^2}{3 \alpha  \left(e^{\sqrt{\frac{2}{3}} \sqrt{\frac{1}{\alpha }} \psi }\right)^2},\,
\quad
\eta = -\frac{2 n e^{-\sqrt{\frac{2}{3}} \sqrt{\frac{1}{\alpha }} \psi }}{3 \alpha } .
\end{eqnarray}
Slow-roll inflation terminates when $\epsilon = 1 $, so the field value at the end of inflation reads
\begin{eqnarray}
\epsilon(\psi_{\rm end}) = 1 = \frac{n^2 e^{-2 \sqrt{\frac{2}{3}} \sqrt{\frac{1}{\alpha }} \psi }}{3 \alpha } \longrightarrow \psi_{\rm end} = \sqrt{\frac{3 \alpha }{8}} \ln \Big(\frac{n^{2}}{3 \alpha }\Big).
\end{eqnarray}
The number of e-foldings for the change of the field $\psi$ from $\psi_{N}$ to $\psi_{\rm end}$ is given by
\begin{eqnarray}
N = \int _{\psi_{\rm end}} ^{\psi_N} d \psi  \frac{V}{dV/d\psi} 
\simeq \frac{3 \alpha  e^{\sqrt{\frac{2}{3}} \sqrt{\frac{1}{\alpha }} \psi_N }}{2 n}- N_{\rm e}
\quad {\rm with}\quad 
N_{\rm e} =\frac{\sqrt{3\alpha}}{2}.
\end{eqnarray}
Therefore, in terms of $N$, the values of $n_{s}$ and $r$ for the small $\alpha$ limit are given by \cite{Ferrara:2013rsa,Kallosh:2013yoa}
\begin{eqnarray}
n_{s}= 1-\frac{2}{ N} ,\,\,\,r = \frac{12 \alpha }{N^{2}}\quad\quad\text{for }\quad \alpha \ll 1 .
\label{smalle}
\end{eqnarray}
It follows from Eqs.~(\ref{attract1-e}) and (\ref{smalle}) that when $\alpha$ is sufficiently large or small, the predictions for the E-model also converge to the attractor given in Eq.~(\ref{attract1}) or the universal attractor given in (\ref{uniattrat}) respectively.
Both T model and E model have the same $\alpha$ attractors because the potentials for the T model and E model have the same asymptotic behavior when $\alpha \ll 1$ and $\alpha \gg 1$. We conclude that the $\alpha$ attractors can be achieved from our multiplicative form models, where the conformal factor can be separated into two parts as in Eq.~(\ref{wmulti}) and $G=g(\phi)W(X)$.
Moreover, the attractors do not depend on the function $W(X)$ in this case. Notice that in this section we just showed that the general scalar-tensor theorieswe considered are equivalent to Einstein gravity with a canonical scalar. Therefore it is clearly possible to choose a potential of any form, including previously studied attractors \cite{Kallosh:2013hoa,Kallosh:20131,Kallosh:20132,Kallosh:2014rga,Galante:2014ifa,Cecotti:2014ipa}. We note that the results present in this section are a slight generalization from Ref.\cite{Kallosh:2014rga}.

\section{Additive form}
\label{add}
Let us now consider the case where $\Omega$ has an additive form, i.e.,
\be
\Omega = w(\phi) + k(\phi) \Xi(X)\,,
\label{omadd}
\ee
where $k(\phi)$ and $\Xi(X)$ are dimensionless.
For this case, Eq.~(\ref{semodv}) becomes
\be
S_E = \int d^4x \sqrt{-\tg}\left[ \frac 12 \tilde{R}
+ \(\frac{G}{\Omega} + \frac 32 \(\frac{w_\phi + k_\phi \Xi}{w + k \Xi}\)^2\) \tilde X
- V_E(\phi)
\right]\,.
\label{semodv-sp1}
\ee
This action can be reduced to Eq.~(\ref{semodv-mt}) if $k(\phi) = k_1 w(\phi)$ and $G = g(\phi) (1 + k_1 \Xi)$,
where $k_1$ is a constant.
Hence, the above action is  a possible generalization of the action in Eq.~(\ref{semodv-mt}).
When $\Omega$ is separated as in Eq.~(\ref{omadd}),
$w(\phi)$ will represent non-minimal coupling and $k(\phi) \Xi(X)$ will represent the non-mimimal kinetic coupling between $\phi$ and gravity.
In analogy to the consideration in section (\ref{multi1}), we set $w(\phi) \equiv 1 + \xi f(\phi)$ and $k(\phi) \equiv \xi_k f_k(\phi)$, where $\xi$ and $\xi_k$ are dimensionless constants while $f(\phi)$ and $f_k(\phi)$ are arbitrary functions.
In the weak non-minimal kinetic coupling limit, i.e., $|k(\phi) \Xi(X)| \ll |w(\phi)|$,
the kinetic terms of $\phi$ in the action (\ref{semodv-sp1}) becomes
\be
 \(\frac{G}{\Omega} + \frac 32 \(\frac{w_\phi + k_\phi \Xi}{w + k \Xi}\)^2\) \tilde X
=  \(\frac{G}{w} + \frac 32 \(\frac{w_\phi }{w}\)^2\) \tilde X\,.
\label{kin1}
\ee
The above kinetic term is similar to that in Eq.~(\ref{semodv-mt}). Hence when the non-minimal kinetic coupling is weak the usual attractor discussed in the previous section can exist.
In the limit where the non-minimal kinetic coupling is strong but the non-minimal coupling is weak, i.e., $|k(\phi) \Xi(X)| \gg |w(\phi)|$ and $w(\phi) \simeq 1$,
the kinetic terms of $\phi$ in the action (\ref{semodv-sp1}) becomes
\be
 \(\frac{G}{\Omega} + \frac 32 \(\frac{w_\phi + k_\phi \Xi}{w + k \Xi}\)^2\) \tilde X
=  \(\frac{G}{k \Xi} + \frac 32 \(\frac{k_\phi }{k}\)^2\) \tilde X\,.
\label{kin2}
\ee
Thus the usual $\alpha$ attractor can exist if $G \equiv g(\phi) \Xi(X)$ and
\be
g(\phi) = \frac{1}{\xi_k}\left(\frac{k_\phi^2 }{k}\right)\,.
\label{sgadd}
\ee
In general when both the non-minimal and non-minimal kinetic couplings are not weak,
the action in Eq.~(\ref{semodv-sp1}) depends on the kinetic term $X$ in the Jordan frame.
The kinetic term $X$ can be eliminated from this action using Eq.~(\ref{xbarx}) to convert $X$ to $\tilde X$ as
\be
\tilde X = \frac{X}{w(\phi) + k(\phi) \Xi(X)}\,,
\quad\Rightarrow\quad
k(\phi) \Xi(X) \tilde X + w(\phi) \tilde X - X = 0\,.
\label{xn}
\ee
For the simplest case where $\Xi(X) \equiv X / \La$ and $\La$ is constant with dimension of mass${}^4$,
the above equation yields
\be
X = \frac{\tilde X w}{1 - k \tilde X / \La}\,.
\label{xadd}
\ee
Therefore
\be
\Omega 
= \frac{w}{1 - k \tilde X /\La}\,.
\label{omegaadd}
\ee
Inserting Eqs.~(\ref{xadd}) and (\ref{omegaadd}) into Eq.~(\ref{semodv-sp1}),
we get
\be
S_E = \int d^4x \sqrt{-\tg}\left\{ \frac 12 \tilde{R}
+ \left[\frac{G}{w}\(1 - k \frac{\tilde{X}}{\La}\) + \frac 3{2 w^2} \(w_\phi + \frac 1{\La} (k_\phi w - w_\phi k)\tilde X\)^2\right] \tilde X
- V_E(\phi)
\right\}\,.
\label{semodv-spx1}
\ee
In principle, the function $G$ can be chosen such that the Lagrangian in the above action is a linear function of $\tilde X$. Consequently we will obtain exactly the same inflationary attractor as discussed in the previous section.
For such choises of $G$, the term $1 + k X/\La$ will appear in the denominator of $G$ in the Jordan frame,
and therefore the  Lagrangian of scalar field $L_2$ does not take a usual form for k-inflation.
In the following consideration, we will see that if $G$ is a polynomial function of $X$,
the action can contain non-linear $\tilde X$-term, and consequently the inflationary predictions have different attractors compared with Eqs.~(\ref{attract1}) and (\ref{uniattrat}).

To perform further analysis, it is necessary to specify forms of $w(\phi), k(\phi)$ and $G(\phi,X)$. For simplicity, one may write these functions in concrete forms,
or keeps one of them generic and then write the other two functions in terms of it.
Here, we consider the second possibility by writing $G(\phi,X) = g(\phi) \gamma(\phi,X)$,
where 
\be
\gamma(\phi,X) \equiv f_0 + f_1(\phi) \frac X\La + f_2(\phi) \frac{X^2}{\La^2} \dots\,,
\label{gamma}
\ee
where all coefficients $f_0, f_1, f_2, \dots$ are dimensionless and $f_0$ is constant.
Similarly to Eq.~(\ref{tc1a}),
$g(\phi)$ is written in terms of $w(\phi)$ as
 \be
g(\phi) = \alpha^2 \frac{ w_\phi^2 }{w}\,,
\label{w2g}
\ee
where $\alpha \equiv 1/(2\sqrt{\xi})$.
For the case where the non-minimal kinetic coupling disappears,
the action in Eq.~(\ref{semodv-spx1}) will not depend on the form of $w(\phi)$ if we can write the action in terms of a new  field variable $\psi$ similar to that in Eq.~(\ref{tca}).
When both the non-minimal and non-minimal kinetic couplings appear in the action, it
is also possible to write the action in Eq.~(\ref{semodv-spx1}) in the form independent of the form of $w(\phi)$ by choosing suitable relation between $k(\phi)$ and $w(\phi)$.
Let us define 
\be
k(\phi) \equiv \kappa \alpha^2 \(\frac{w_\phi}{w}\)^2\,,
\label{k2wp}
\ee
where $\kappa$ is constant,
so that the action (\ref{semodv-spx1}) can be written as
\be
S_E = \int d^4x \sqrt{-\tg}\left\{ \frac 12 \tilde{R}
+ \left[\gamma(\phi, X)\(1 - \kappa \frac{X_\psi}{\La}\) + \frac 3{2\alpha^2}
\(1 - 3 \kappa \frac{X_\psi}{\La}
+ 2\kappa\frac{w_{\phi\phi}w}{w_\phi^2}\frac{X_\psi}{\La}\)^2
\right] X_\psi
- V(\psi)
\right\}\,,
\label{act-psi-add}
\ee
where $X_\psi \equiv -\tilde g^{\mu\nu} \partial_\mu\psi \partial_\nu \psi / 2$ and $\psi$ is defined via
\be
w(\phi) =  \exp\(\frac{\psi}{\alpha}\)\,.
\label{w2psiadd}
\ee
It can be seen that the action in Eq.~(\ref{act-psi-add}) still depends on the form of $w(\phi)$ unless $w_{\phi\phi}w / w_{\phi}^2$ is constant.
The constancy of the ratio $w_{\phi\phi}w / w_{\phi}^2$ is possible for various forms of $w(\phi)$, for examples, $w \sim e^{\xi \phi}, w \sim \cosh(\xi \phi)$, etc, and also $w = (1 + \xi \phi^p)$ with large coupling constant $\xi$.
Moreover, this ratio is expected to be nearly constant for arbitrary form of $w(\phi)$ when $\phi$ slowly varies with time.
Hence, it is reasonable to suppose that the ratio $w_{\phi\phi}w/w_\phi^2$ is constant and can be quantified by
\be
\frac{w_{\phi\phi}w}{w_\phi^2} = \lambda\,,
\label{wratio}
\ee
where $\lambda$ is a constant parameter.
Inserting the above relation into Eq.~(\ref{act-psi-add}),
we get
\be
S_E = \int d^4x \sqrt{-\tg}\left\{ \frac 12 \tilde{R}
+ \left[\gamma(\phi, X)\(1 - \kappa \frac{X_\psi}{\La}\) + \frac 3{2\alpha^2}
\(1 - 3 \kappa \frac{X_\psi}{\La}
+ 2\kappa\lambda\frac{X_\psi}{\La}\)^2
\right] X_\psi
- V(\psi)
\right\}\,,
\label{act-psi-add1}
\ee
Firstly, we consider the case where $\gamma(\phi, X)$ is constant, but not equal to $- 3/(2\alpha^2)$.
For this case, the slow-roll evolution of $\psi$ during inflation suggests that the $\tilde X^2$-term and $\tilde X^3$-term in the action (\ref{act-psi-add1}) can be neglected.
Consequently, the theoretical predictions from inflationary model described by the action (\ref{act-psi-add1})  obeys the attractor in Eqs.~(\ref{attract1}) and (\ref{uniattrat}) under suitable redefinition of parameter $\alpha$.

For the case of $\gamma(\phi, X) = - 3/(2\alpha^2)$,
the linear $X_\psi$-term in the action (\ref{act-psi-add1}) disappears and then under the slow roll approximation the kinetic term of $\psi$ is proportional to $X_\psi^2$. Hence the action becomes
\be
S_E \simeq \int d^4x \sqrt{-\tg}\left\{ \frac 12 \tilde{R}
+ \frac{X_\psi^2}{\La_2}
- V(\psi)
\right\}\,,
\label{act-psi-add2-x2}
\ee
where 
\be
\La_2 \equiv
\left[\frac{3\kappa}{2\alpha^2}\(4\lambda - 5\)\right]^{-1} \La\,.
\ee
The observable quantities for this case will be discussed in the subsequent studies. Another interesting form of $\gamma(X)$ is the form where $\gamma(X)$ is a linear function of $X$ as
\be
\gamma(\phi, X) = 
\frac 3{2\alpha^2}\left[\(5 - 4 \lambda\) \frac{X}{\La} \frac{k(\phi)}{w(\phi)} - 1\right]\,.
\label{gx3}
\ee
The above equation can be written in terms of $X_\psi$ as
\be
\gamma = 
\frac 3{2\alpha^2}\left[\(5 - 4 \lambda\) \kappa \frac{X_\psi}{\La - \kappa X_\psi} - 1\right]\,.
\label{gbx3}
\ee
Inserting this relation into Eq.~(\ref{act-psi-add1}),
we get
\be
S_E = \int d^4x \sqrt{-\tg}\left\{ \frac 12 \tilde{R}
+ \frac{X_\psi^3}{\La_3^2}
- V(\psi)
\right\}\,,
\label{act-psi-add2-x3}
\ee
where 
\be
\La_3 \equiv
\left[\frac{3\kappa^2}{2\alpha^2}\(2\lambda - 3\)^2 \right]^{-1} \La\,.
\ee
The observable predictions from inflationary models described by the actions in Eqs.~(\ref{act-psi-add2-x2}) and (\ref{act-psi-add2-x3}) have different attractors from those given in Eqs.~(\ref{attract1}) and (\ref{uniattrat}).
We will study this attractor in the following considerations.
In general, it is also possible to set $\gamma(\phi,X) \propto X^m$ where $m \geq 2$.
Nevertheless, it leads to the term proportional to $X_\phi^{m+2}$ which is negligible in the slow roll limit.
To compute the observable quantities,
we use the slow-roll approximation in which the evolution equations derived from the actions (\ref{act-psi-add2-x2}) and (\ref{act-psi-add2-x3}) can be written as
\be
H^2 \simeq \frac 13 V(\psi)\,,
\qquad\mbox{and}\qquad
(\psi')^{2q-1} = - A_q \La_q^{q-1} \frac{1}{V^q}\frac{d V}{d \psi}\,,
\label{eom-slow}
\ee
where $H$ is a Hubble parameter, a prime denotes derivative with respect to $\ln a$, $a$ is a cosmic scale factor, $A_q \equiv 6^{q - 1}/q$,
and $q = 2, 3$ for $X_\phi^2$ and $X_\phi^3$ models respectively.
Since the form of the equation of motion for scalar field $\psi$ is different from that for the usual canonical normalized field,
we have to  compute the slow roll parameter $\epsilon$ and number of e-folding $N$ from their definitions:
\be
\epsilon \equiv - \frac{\dot H}{H^2}\,,
\qquad\mbox{and}\qquad
N \equiv \int H dt\,.
\label{full-epn}
\ee
Using Eq.~(\ref{eom-slow}),
the relations in Eq.~(\ref{full-epn}) can be written as
\be
\epsilon = 
\frac{A_q^{1/(2q-1)}}{2} \La_q^{(q-1)/(2q-1)}\frac{1}{V^{(3q-1)/(2q-1)}}\(\frac{dV}{d \psi}\)^{2q/(2q-1)},
\label{epsiadd}
\ee
and 
\be
N = A_q^{1/(1-2q)}\La_q^{(1-q)/(2q-1)} \int _{\psi_{\rm end}} ^{\psi_N} d \psi  \frac{V^{q/(2q-1)}}{(dV/d\psi)^{1/(2q-1)}}\,.
\label{nadd}
\ee
For the inflaton with non-canonical kinetic terms,
the spectral index and the tensor-to-scalar ratio are given by \cite{Garriga:99, Lorenz:08}
\ba
n_{s} &=& 1 - 2 \epsilon 
- \psi' \frac{d}{d\psi}\(\ln \epsilon\)
- \psi' \frac{d}{d\psi}\(\ln  c_s^2\)\,,
\label{sindex}\\
r &=& 16 c_{s} \epsilon\,,
\label{ratio}
\ea
where $c_s \equiv \sqrt{(\partial P/\partial X_\psi)/ (\partial \rho/\partial X_\psi)} = 1 / \sqrt{2q-1}$ is the propagation speed of the scalar perturbations,
$P = X_\psi^q/\La^{q-1} - V(\psi)$ is the pressure of $\psi$,
and $\rho \equiv (2q-1) X_\psi^q/\La^{q-1} + V(\psi)$ is the energy density of $\psi$.

\subsection{Hyperbolic tangent potential (T-model)}
To obtain the potential for the T-model,
we set $V_{E}(\phi)=V_{0} \La_q \left[\Big(w - 1\Big)/\Big(w + 1\Big)\right]^{n}$ to obtain
\be
V(\psi) = V_{0} \La_q \left[\tanh\Big(\frac{\psi}{2 \alpha}\Big)\right]^{n},
\label{t-add}
\ee
where $V_0$ is dimensionless constant which is supposed to be order of unity.
In the following  analytical analysis,
we will restrict ourselves to the case $n = 2$ in which the analytical expressions for $n_s$ and $r$ in terms of the number of e-folding can be straightforwardly obtained.
This restriction will be relaxed when numerical analysis is performed in section \ref{obs}.
Substituting this potential into Eqs.~(\ref{epsiadd}) and (\ref{nadd}) and then setting $n = 2$,
we get
\ba
\epsilon &=& 
\frac 12 \(\frac{A_q}{V_0^{(q-1)}\alpha^{2q}}\)^{1/(2q-1)}
\left[\sinh^2\(\frac\psi{2\alpha}\)\cosh^{2/(2q -1)}\(\frac\psi{2\alpha}\)\right]^{-1}\,,
\label{epsi-add2}\\
N &=&  (2q-1)\left[\frac{V_0^{q-1} \alpha^{2q}}{A_q}
\cosh^2\(\frac\psi{2\alpha}\)\right]^{1/(2q-1)}\biggr\rvert_{\psi_{\rm end}}^{\psi_N}\,.
\label{n-add02}
\ea
In this case, the values of the field $\psi$ at the end of inflation cannot be computed analytically from the relation $\epsilon = 1$.
Hence, we define
\be
N_e \equiv  (2q-1)\left[\frac{V_0^{q-1} \alpha^{2q}}{A_q}
\cosh^2\(\frac{\psi_{\rm end}}{2\alpha}\)\right]^{1/(2q-1)}\,,
\label{ne2}
\ee
so that  Eq.~(\ref{n-add02}) can be written as
\be
N + N_e =  A \cosh^{2/(2q-1)}\(\frac{\psi_N}{2\alpha}\)\,,
\label{n-add2}
\ee
where
\be
A \equiv 
(2q-1)\left[\frac{V_0^{q-1} \alpha^{2q}}{A_q}\right]^{1/(2q-1)}\,.
\label{defa}
\ee
Eq.~(\ref{epsi-add2}) can be written in terms of the number of e-folding using Eq.~(\ref{n-add2}) as
\be
\epsilon_N =
\frac{(2q-1) A^(2q-1)}{2\(N + N_e\) \left(\(N + N_e\)^{2q-1} - A^{2q-1}\right)}\,.
\label{epsi-term-nad2}
\ee
Using Eqs.~(\ref{epsi-add2}) and (\ref{n-add2}),
Eqs.~(\ref{sindex}) and (\ref{ratio}) can be written as
\ba
n_s &=& 1 - \frac{2q}{N + N_e} 
- \frac{(4q-2) A^(2q-1)}{\(N + N_e\) \left[\(N + N_e\)^{2q-1} - A^{2q-1}\right]}\,,
\label{ns-add2}\\
r &=& 16 c_s \frac{(2q-1) A^(2q-1)}{2\(N + N_e\) \left[\(N + N_e\)^{2q-1} - A^{2q-1}\right]}\,.
\label{r-add2}
\ea
For $\alpha \gg 1$ or equivalently in the weak coupling limit $\xi \ll 1$,
the condition $\epsilon = 1$ at the end of inflation yields $\psi_{\rm end} \simeq \alpha \sqrt{(4q - 2) / A}$,
and therefore 
\be
N_e \simeq \(1 + \frac 1{2A}\) A \simeq A + \frac 12\,.
\label{ne-add2-ala}
\ee
Substituting the above relation into Eqs.~(\ref{r-add2}) and (\ref{ns-add2}),
we get
\be
n_s \simeq 1 - \frac{8q - 4}{(4q - 2)N + (2q-1)}\,,
\qquad
r \simeq  16 \frac{\sqrt{2q -1}}{(4q - 2) N + (2q-1)}\,.
\label{rns-add2-laa}
\ee
Interestingly, the theoretical predictions in this limit do not depend on $\kappa$ which controls relative strength between the non-minimal and non-minimal kinetic couplings.
In contrast, for $\alpha \ll 1$ or equivalently strong coupling $\xi \gg 1$ limit,
$\epsilon = 1$ gives $e^{2q\psi_{\rm end}/ ((2q-1) \alpha)} \simeq 2^{4q/(2q-1)} (2q-1) / (2 A)$,
so that 
\be
N_e \simeq \(\frac 52\)^{1/2q} A^{(2q-1)/2q} \propto \alpha\,,
\label{ne-add2-sma}
\ee
and consequently
\be
n_s \simeq   1 - \frac{2q}{N} + {\cal O}\(\frac \alpha{N^2}\)\,,
\qquad
r \simeq \frac{8}{3} (2q-1)^{2q - 1/2} \frac{V_{0}^{(q-1)}\alpha^{2q}}{N^{2q}} + {\cal O}\(\frac{\alpha^{2q+1}}{N^{2q+1}}\)\,.
\label{nsr-add2-sma}
\ee
Again, the inflationary predictions do not depend on $\kappa$.
Note that $r$ is independent of $\kappa$ because the coefficient of the potential defined in Eq.~(\ref{t-add}) is in the form of $V_0 \La_q$ with constant $V_0$.
Instead of setting $V_0$ to be constant, if we set $\tilde{V}_0 \equiv V_0 \La_q$ to be constant independent of $\kappa$ and $\alpha$ ,
the expression for $r$ will depend on $\kappa$ when $V_0$ is replaced by $\tilde{V}_0$.

It follows from Eqs.~(\ref{rns-add2-laa}) and (\ref{nsr-add2-sma})  that at $\alpha \gg 1$ and $\alpha \ll 1$ limits,
the expressions for observable quantities, $n_s$ and $r$, converge to the forms that are similar to Eqs.~(\ref{attract1}) and (\ref{uniattrat}) up to some constant factors.
The existence of these convergences does not depend on the form of $w(\phi)$ but of course depends on the relation among $w(\phi)$, $k(\phi)$ and $G(\phi,X)$.
Moreover, Eqs.~(\ref{rns-add2-laa}) and (\ref{nsr-add2-sma}) are computed in the large and small $\alpha$ limits,
in which various potentials take similar formes, especially the potentials for the T model and E model described  in the previous sections.
Hence,  the convergence of the observable quantities to Eqs.~(\ref{rns-add2-laa}) or (\ref{nsr-add2-sma}) at asymptotic value of $\alpha$
can imply the inflationary attractor.

In more general cases where $n$ is not restricted to be two,
the number of e-folding will depend on Hypergeometric functions so that it is not possible to write $\epsilon$ in terms of the number of e-folding. In this situation, it is difficult to write analytic expressions for $n_s$ and $r$ in terms of the number of e-folding.

\subsection{Theoretical predictions for large and small $\alpha$ limits}
\label{obs}

As mentioned previously, the potentials of the T model and E model have the same asymptotic behavior in the large and small $\alpha$ limits. Since the inflationary attractors are characterized by these asymptotic behaviors,
we investigate in this section inflationary predictions for the models described by Eq.~(\ref{eom-slow}) in the large and small $\alpha$ limits instead of repleting the calculations in the previous section for the E model.

In the limit $\alpha \gg 1$,
the potential in Eq.~(\ref{t-add}) becomes 
\be
V(\psi) \simeq V_0 \La_q \(\frac \psi{2 \alpha}\)^n\,.
\label{lalpot}
\ee
Replacing the potential in Eq.~(\ref{t-add}) by this approximated potential,
it can be shown that $\epsilon$ and $N$ are given by
\ba
\epsilon &=& 
\frac 12 \(\frac{A_q n^{2q}\psi^{n -2q - nq}}{V_0^{(q-1)} (2\alpha)^{n - nq}}\)^{1/(2q-1)}\,,
\label{epsi-addas}\\
N &=& 
\frac{2q-1}{2q + nq - n}\left[\frac{V_0^{q-1} (2\alpha)^{n-nq}}{n A_q \psi_N^{n -2q - nq}}\right]^{1/(2q-1)}
- \frac n2 \frac{2q-1}{2q + nq - n}\,.
\label{n-addas}
\ea
Combinding the above two equations,
we can write $\epsilon$ in terms of the number of e-folding as
\be
\epsilon_N = 
\frac n2 \frac{2q-1}{2q + nq - n}\left[N + \frac n2 \frac{2q-1}{2q + nq - n}\right]^{-1}\,.
\label{epsi-n-addlal}
\ee
Inserting the above results into Eqs.~(\ref{sindex}) and (\ref{ratio}),
we get
\be
n_s = 1 - 2 \frac{2q + 3nq - 2n}{2(2q + nq - n)N + n(2q-1)}\,,
\qquad
r = \frac{16}{\sqrt{2q-1}} \frac{2qn -n}{2 (2q + nq - n) N + n(2q-1)}\,.
\label{nsr-genlal}
\ee
In the limit $\alpha \ll1 $,
the potential in Eq.~(\ref{t-add}) becomes
\be
V(\psi) \simeq V_0\La_q (1 - 2 n e^{-\psi/\alpha})\,,
\label{salpot}
\ee
and therefore we have
\begin{eqnarray}
\epsilon 
&=& \frac 12 B \left[\frac{e^{- 2q \psi/\alpha}}{\(1 - 2n e^{-\psi/\alpha}\)^{3q-1}}\right]^{1/(2q-1)}
\simeq \frac 12  B e^{- 2q \psi/(2q-1)\alpha}\,,
\label{epsi-addas1}\\
N &=& \frac{2n}{B}\((2q-1) - \frac{n q}{1-q} e^{-\psi/\alpha}\) e^{\psi/(2q-1)\alpha}\biggr\rvert_{\psi_{\rm end}}^{\psi_N}\nonumber\\
&\simeq& \frac{2n}{B} \(2q-1\)e^{\psi_N/(2q-1)\alpha} - (2q-1) \alpha \(\frac{V_0^{q-1}}{2A_q}\)^{1/2q}\,,
\label{n-addas1}
\end{eqnarray}
where 
\be
B \equiv \left[\frac{A_q (2n)^{2q}}{V_0^{q-1} \alpha^{2q}}\right]^{1/(2q-1)}\,.
\ee
 In terms of the number of e-folding, $\epsilon$ can be written as
\be
\epsilon_N \simeq
\frac 12 (2q-1)^{2q} \alpha^{2q} \frac{V_0^{q-1}}{A_q}\left[N + (2q-1) \alpha\(\frac{V_0^{q-1}}{2A_q}\)^{1/2q}\right]^{-2q}\,.
\label{epsi-n-addsal}
\ee
Hence, for this case, Eqs.~(\ref{sindex}) and (\ref{ratio}) yield
\be
n_{s} \simeq 1 - \frac{2q}{N}\,,
\qquad
r \simeq \(2q-1\)^{2q} \frac{8}{\sqrt{2q-1}} \frac{V_0^{q-1}}{A_q}\frac{\alpha^{2q}}{N^{2q}}\,.
\label{nsr-gensal}
\ee
Eqs.~(\ref{nsr-genlal}) and (\ref{nsr-gensal}) are the generalization of the attractor in Eqs.~(\ref{rns-add2-laa}) and (\ref{nsr-add2-sma}).
These equations will become the  attractors in Eqs.~(\ref{attract1}) and (\ref{uniattrat}) if $q = 1$ for suitable redefinition of $\alpha$.
We will see in the numerical investigation that due to the factor $2q$ in the expression for $n_s$ in Eq.~(\ref{nsr-gensal}),
the value of $n_s$ in small $\alpha$ limit is less than the observational bound and values from universal attractor at large number of e-folding, e.g., at $N = 60$.
This puts a tight constraint on the inflationary attractor in the strong coupling $\xi \gg 1$ limit for the models where $q > 1$.
\begin{figure}[H]
\begin{center}		
\includegraphics[width=0.47\linewidth]{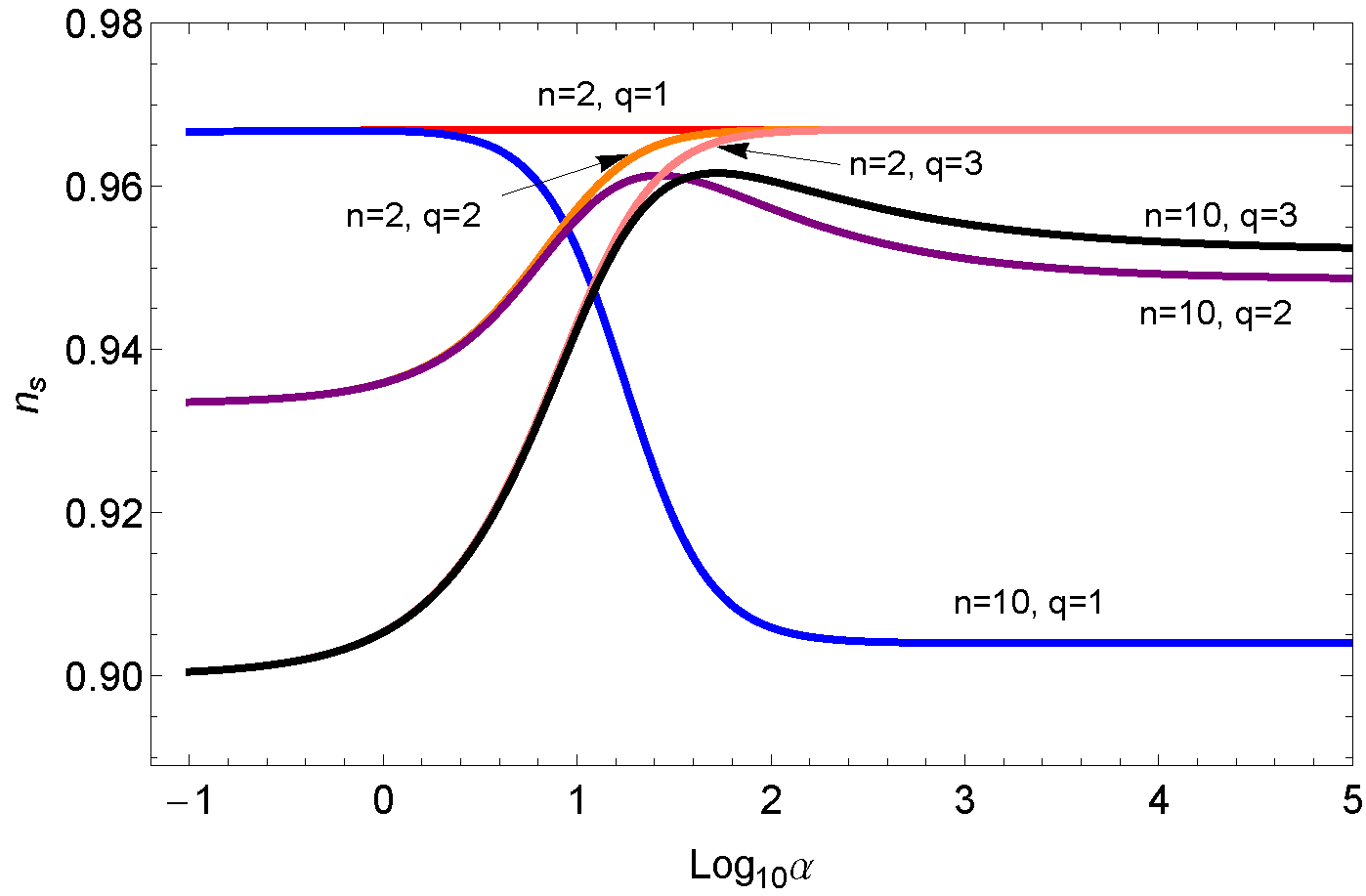}
\includegraphics[width=0.47\linewidth]{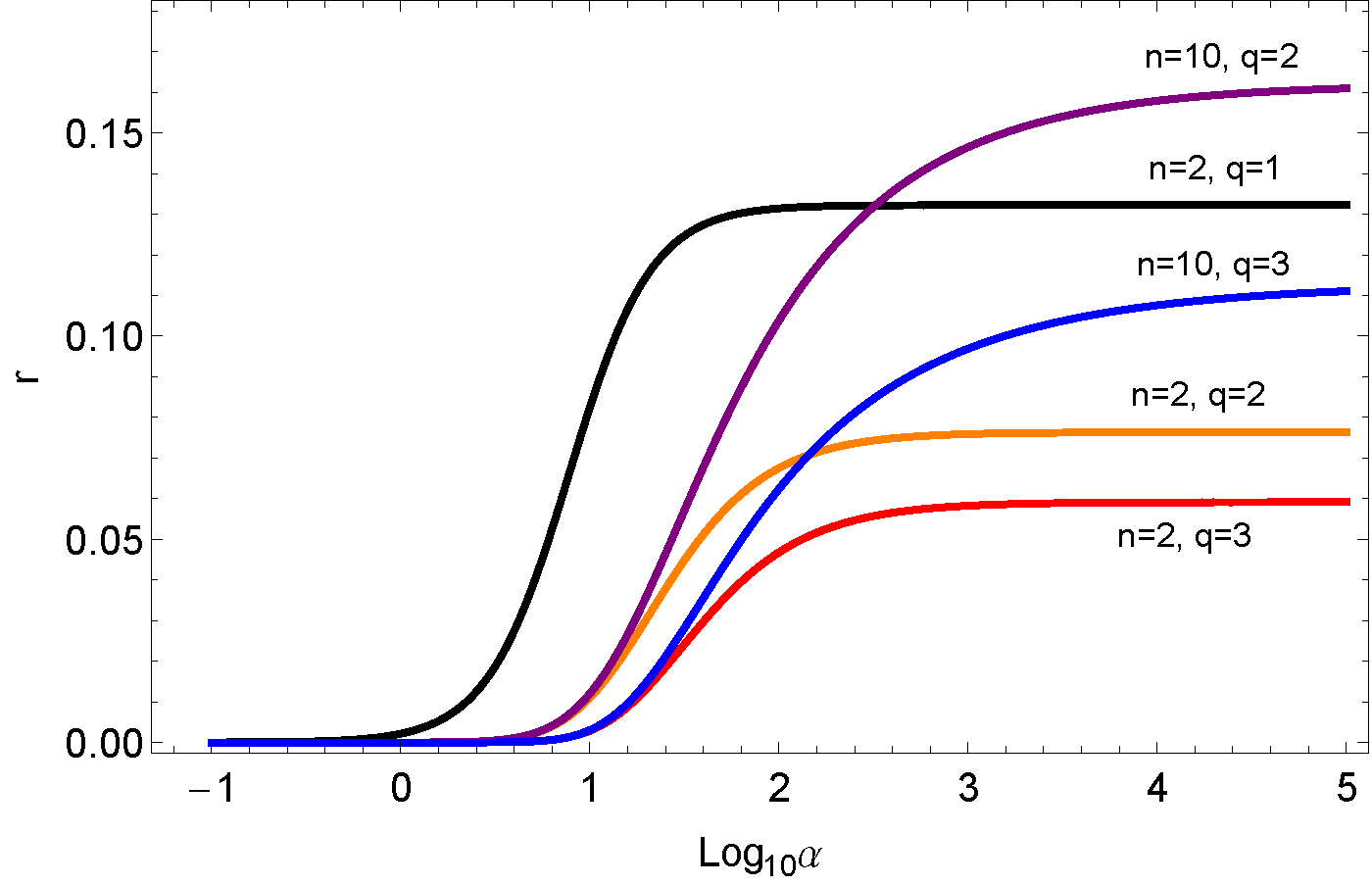}
\caption{\label{figlrns}
The plots show how $n_{s}$ and $r$ evolve with the changing of $\alpha$. In all plots, $n_s$ and $r$ are evaluated at $N = 60$.}
\end{center}
\end{figure}
In order to compute the observable  quantities from the inflationary models whose dynamics are governed by evolution equations in Eq.~(\ref{eom-slow}) and potential is given by Eq.~(\ref{t-add}),
we integrate Eq.~(\ref{eom-slow}) and compute the observable quantities numerically for various values of $q, n$ and $\alpha$.

In Fig.\ref{figlrns}, we plot the predictions of the model in the $n_{s}-\log(\alpha)$ and $r-\log(\alpha)$  plane for various values of the parameters $n$ and $q$. From Fig.\ref{figlrns}, we discover that our results for $\alpha < O(10)$ with any values of $n$ and $q$ show an attractor behavior, but with only $q=1$ display an universal attractor given in Eq.(\ref{uniattrat}). From our definition of $\alpha=1/(2\sqrt{\xi})$, we see that the attractor can be achieved when $\xi > O(10^{-3})$ which is in agreement with Ref.\cite{Kallosh:20131}. In addition, from Fig.\ref{figlrns}, in case of large $\alpha$ the attractor can be achieved when $\xi < O(10^{-4})$.

\subsection{Contact with recent Planck data}
\label{planck}
\begin{figure}[H]
\begin{center}		
\includegraphics[width=0.47\linewidth]{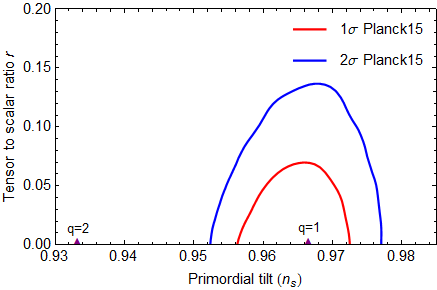}
\caption{\label{figsm}
 In case of small values of $\alpha$, we compare the theoretical predictions in the $(n_{s}-r)$ plane for small $\alpha$ with Planck 2015
results for TT, TE, EE, +lowP and assuming $\Lambda$CDM + r \cite{Ade:2015lrj}.}
\end{center}
\end{figure}

\begin{figure}[H]
\begin{center}		
\includegraphics[width=0.47\linewidth]{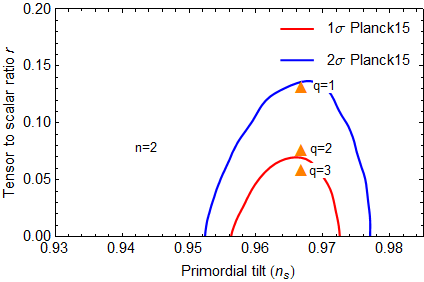}
\includegraphics[width=0.47\linewidth]{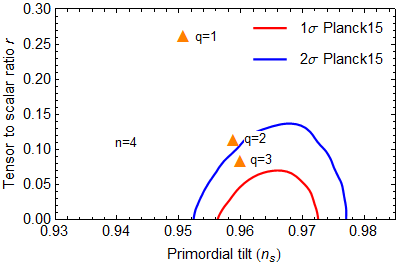}
\includegraphics[width=0.47\linewidth]{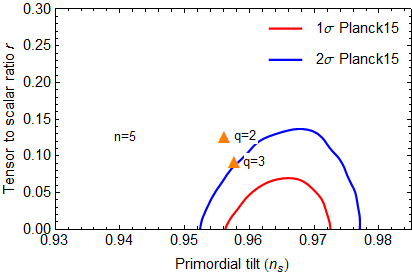}
\caption{\label{figlal}
 In case of large values of $\alpha$, we compare the theoretical predictions in the $(n_{s}-r)$ plane for large $\alpha$ with Planck015
results for TT, TE, EE, +lowP and assuming $\Lambda$CDM + r \cite{Ade:2015lrj}.}
\end{center}
\end{figure}

In this section, we compare our results in Eqs.~(\ref{nsr-genlal}) and (\ref{nsr-gensal}) with Planck 2015 data. Note once that the potentials of the T model and E model have the same asymptotic behavior in the large and small $\alpha$ limits. In the small $\alpha$ limit, we compared our results with the Planck 2015 measurement by placing the predictions in the $(n_{s}-r)$ plane with different values of $q$ while kept $N=60$, illustrated in Fig.\ref{figsm}. We notice that with $q=1$ our results lie within $2\sigma$\,C.L. of Planck 2015 contours. However, when $q>1$ the results are far outside $2\sigma$\,C.L. of Planck 2015 contours.  In addition, from the right panel of Fig.\ref{figlrns}, for various values of $n$ and $q$ at strong coupling limit our model provides $r<0.064$ in precise agreement with the improved value recently reported in \cite{Akrami:2018odb}.
However, in the large $\alpha$ limits, with $n=2$ our results lie within $1\sigma$\,C.L. of Planck 2015 contours for $q=1\,\&\,2$, while within $2\sigma$\,C.L. of Planck 2015 contours for $q=3$, illustrated in the upper-left panel of Fig.\ref{figlal}. Moreover, our results lie far outside $2\sigma$\,C.L. of Planck 2015 when $q=1, n=4$, but lie within $1\sigma$\,C.L. of Planck 2015 when $q=3$ with $n=4$, displayed in the upper-right panel of Fig.\ref{figlal}.
 In the lower-panel of Fig.\ref{figlal} with $n=5$ we observe that when $q=2$ the results lie outside $1\sigma$\,C.L. of Planck 2015 contours, while $q=3$ our results lie inside $1\sigma$\,C.L. of Planck 2015. In addition, we can deduce that when $q>3$ in this case the results lie well within $2\sigma$\,C.L. of Planck 2015 contours. Interestingly we conclude that the greater values of $q$ take, the better results lie well within $2\sigma$\,C.L. of Planck 2015 contours for any $n$.

\section{conclusion}
Among viable inflationary models, the $\alpha$ attractors, in light of the presently existing CMB data, has received particular attention. In the present work, we investigated the inflationary attractors in models of inflation inspired by general conformal transformation  of general scalar-tensor theories to the Einstein frame. Since the coefficient of the conformal transformation in our study depends on both the scalar field and its kinetic term,
the non-minimal  coupling in the presence of both the field and its kinetic term can appear in the action in the Jordan frame.
This action presents a subset of the class I of the DHOST theories, and therefore the theories associated to this action free from the Ostrogradski instability. 

In our analysis, we concentrated on the inflationary models in the Einstein frame.
Nevertheless, according to a brief consideration in Sec.~(\ref{sec1}),
the number of e-folding in the Einstein frame is approximately equal to that in the Jordan frame in the strong $\xi \gg 1$ and weak $\xi \ll 1$ coupling limits.
Hence, the observational quantities  in terms of number of e-folding are approximately frame-invariant,
and consequently the inflationary attractors in the Einstein frame should imply the existence of the same attractors in the Jordan frame in the strong and weak coupling limits.

We considered the two-case scenarios. We first concentrated on the multiplicative form of the generalized conformal factor, i.e. $\Omega=\Omega(X,\phi)=w(\phi)W(X)$. 
The action in the Einstein frame does not depend on $W(X)$ if the coefficient $G(\phi, X)$ of the kinetic term of $\phi$ in the Jordan frame takes the form $G(\phi, X) = g(\phi)W(X)$.
Based on this setting, we have proposed the models which are designed specifically to be the T-model and E-model actions. The main finding from the multiplicative form model is that the usual $\alpha$ attractors can be achieved from models constructed by the generalized conformal transformation. 
 From our definition of $\alpha$ in terms of the coupling constant $\xi$,
we have $\alpha \to \infty$ in the weak coupling limit and $\alpha \to 1$ in the strong coupling limit.
In the strong coupling limit, the predictions converge to the universal attractor regime in Eq.~(\ref{uniattrat}) \cite{Kallosh:2013yoa,Ferrara:2013rsa, Galante:2014ifa} which corresponds to the part
of the $n_{s}- r$ plane favored by the Planck data \cite{Ade:2013zuv}.
For small coupling limit, the predictions  converge to  Eq.~(\ref{attract1}) if $n$ is replaced by $2n$.

In addition, we have chosen the additive form of the generalized conformal factor, i.e. $\Omega=\Omega(X,\phi)=w(\phi)+k(\phi) X$. We also compute the cosmological observables for T-model potential. We have found that in our choice of the relation among the functions of the coefficients,
the inflationary predictions do not depend on both $w(\phi)$ and relative strength between the non-minimal kinetic and usual non-minimal couplings.
However, in some choices of the relation among the functions of the coefficients,
the kinetic term of the redefined field that governs dynamics of inflation takes a non-linear form, e.g., $X_\psi^2$ and $X_\psi^3$.
In these situations, the inflationary predictions converge to new attractors given by Eqs.~(\ref{nsr-genlal}) and (\ref{nsr-gensal})
in the weak and strong coupling limits respectively.
For the additive form of the conformal factor, the parameter $\alpha$ is defined such that the weak and strong coupling limits are equivalent to large and small $\alpha$ respectively. From our numerical calculation,  we discovered that the attractor can be achieved for the strong coupling limit and the weak one when $\xi > O(10^{-3})$ and $\xi < O(10^{-4})$, respectively.

We confronted the obtained results of the  cosmological observables with recent Planck 2015 data
More concretely, in the small $\alpha$ limit, we compared our results given in Eq.(\ref{nsr-gensal}) with the Planck 2015 measurement by placing the predictions in the $(n_{s}-r)$ plane with different values of $q$ while kept $N=60$, illustrated in Fig.\ref{figsm}. We notice that with $q=1$ our results lie within $1\sigma$\,C.L. of Planck 2015 contours. However, when $q>1$ the results are not satisfied the observational bound of the Planck 2015 contours. However, in the large $\alpha$ limits given in Eq.(\ref{nsr-genlal}), with $n=2$ our results lie within $1\sigma$\,C.L. of Planck 2015 contours for $q=1\,\&\,2$, while within $2\sigma$\,C.L. of Planck 2015 contours for $q=3$, illustrated in the upper-left panel of Fig.\ref{figlal}. Moreover, our results lie far outside $2\sigma$\,C.L. of Planck 2015 when $q=1, n=4$, but lie within $1\sigma$\,C.L. of Planck 2015 when $q=3$ with $n=4$, displayed in the upper-right panel of Fig.\ref{figlal}. Notice that the greater values of $q$ take, the better results lie well within $2\sigma$\,C.L. of Planck 2015 contours for $n=4$. In the lower-panel of Fig.\ref{figlal} with $n=5$ we observe that when $q=2$ the results lie outside $1\sigma$\,C.L. of Planck 2015 contours, while $q=3$ our results lie inside $1\sigma$\,C.L. of Planck 2015. In addition, we can deduce that when $q>3$ in this case the results lie well within $2\sigma$\,C.L. of Planck 2015 contours.

Notice that we started in Sec.(\ref{sec1}) by considering a generic form of coefficients for conformal transformation, and  restricted our subsequent discussions focusing two-case scenarios taking the multiplicative and additive forms of the conformal coefficient for simplicity. More precisely, the multiplicative form model  is chosen in such a way that the standard $\alpha$ attractors can  be recovered in the models where conformal coefficient depends on the kinetic term of scalar field. Moreover, the new inflationary attractors can be achieved by choosing the additive form model which can be viewed as the extension of multiplicative form model.
However, the additive form of the conformal coefficient is restricted such that the exact relation between the kinetic terms of scalar field in the Jordan and  Einstein frames can be obtained. In the simplest case, this relation is presented in Eq.~(\ref{xadd}). Hence, there should be inflationary attractors other than those present in this work if this restriction is relaxed. We will leave this interesting topic for our future investigation.

\subsection*{Acknowledgement}
This work is financially supported by the Institute for the Promotion of Teaching Science and Technology (IPST) under the project of the \lq\lq Research Fund for DPST Graduate with First Placement\rq\rq\,, under Grant No.\,033/2557 and partially supported by Thailand Center of Excellence in Physics (ThEP) with Grant No.\,ThEP-61-PHY-NU1. Valuable comments
and intuitive suggestions from the referee are also acknowledged.


\begin{thebibliography}{50}

\bibitem{Bezrukov:2007ep} 
  F.~L.~Bezrukov and M.~Shaposhnikov,
  Phys.\ Lett.\ B {\bf 659}, 703 (2008)

\bibitem{Barvinsky:2008ia} 
  A.~O.~Barvinsky, A.~Y.~Kamenshchik and A.~A.~Starobinsky,
  JCAP {\bf 0811}, 021 (2008)

\bibitem{Bezrukov:2008ut} 
  F.~Bezrukov, D.~Gorbunov and M.~Shaposhnikov,
  JCAP {\bf 0906}, 029 (2009)

\bibitem{Bezrukov:2008ej} 
  F.~L.~Bezrukov, A.~Magnin and M.~Shaposhnikov,
  Phys.\ Lett.\ B {\bf 675}, 88 (2009)

\bibitem{Bezrukov:2009db} 
  F.~Bezrukov and M.~Shaposhnikov,
  JHEP {\bf 0907}, 089 (2009)

\bibitem{Barbon:2009ya} 
  J.~L.~F.~Barbon and J.~R.~Espinosa,
  Phys.\ Rev.\ D {\bf 79}, 081302 (2009)


\bibitem{Evans:2010tf} 
  N.~Evans, J.~French and K.~y.~Kim,
  JHEP {\bf 1011}, 145 (2010)

\bibitem{Channuie:2011rq} 
  P.~Channuie, J.~J.~Joergensen and F.~Sannino,
  JCAP {\bf 1105}, 007 (2011)

\bibitem{Bezrukov:2011mv} 
  F.~Bezrukov, P.~Channuie, J.~J.~Joergensen and F.~Sannino,
  Phys.\ Rev.\ D {\bf 86}, 063513 (2012)

\bibitem{Channuie:2012bv} 
  P.~Channuie, J.~J.~Jorgensen and F.~Sannino,
  Phys.\ Rev.\ D {\bf 86}, 125035 (2012)

\bibitem{Channuie:2016iyy} 
  P.~Channuie and C.~Xiong,
  Phys.\ Rev.\ D {\bf 95}, no. 4, 043521 (2017)

\bibitem{Samart:2018ucu} 
  D.~Samart and P.~Channuie,
  arXiv:1807.10724 [hep-th]


\bibitem{Kallosh:2013hoa} 
  R.~Kallosh and A.~Linde,
  JCAP {\bf 1307}, 002 (2013)

\bibitem{Kallosh:20131}
R.~Kallosh and A.~Linde,
JCAP {\bf 1310}, 033 (2013)

\bibitem{Kallosh:20132}
R.~Kallosh, A.~Linde, D.~Roest,
Phys.\ Rev.\ Lett.\ {\bf 112}, 011303 (2014)

\bibitem{Kallosh:2014rga} 
  R.~Kallosh, A.~Linde and D.~Roest,
  JHEP {\bf 1408}, 052 (2014)

\bibitem{Galante:2014ifa} 
  M.~Galante, R.~Kallosh, A.~Linde and D.~Roest,
  Phys.\ Rev.\ Lett.\  {\bf 114}, no. 14, 141302 (2015)

\bibitem{Cecotti:2014ipa} 
  S.~Cecotti and R.~Kallosh,
  JHEP {\bf 1405}, 114 (2014)

\bibitem{Yi:2016jqr} 
  Z.~Yi and Y.~Gong,
  Phys.\ Rev.\ D {\bf 94}, no. 10, 103527 (2016)

\bibitem{Carrasco:2015rva} 
  J.~J.~M.~Carrasco, R.~Kallosh and A.~Linde,
  Phys.\ Rev.\ D {\bf 92}, no. 6, 063519 (2015)

\bibitem{Carrasco:2015uma} 
  J.~J.~M.~Carrasco, R.~Kallosh, A.~Linde and D.~Roest,
  Phys.\ Rev.\ D {\bf 92}, no. 4, 041301 (2015)


\bibitem{Ade:2015lrj} 
  P.~A.~R.~Ade {\it et al.} [Planck Collaboration],
  Astron.\ Astrophys.\  {\bf 594}, A20 (2016)

\bibitem{Akrami:2018odb}
  Y.~Akrami {\it et al.} [Planck Collaboration],
  arXiv:1807.06211 [astro-ph.CO].

\bibitem{Zumalaca}
  M.~Zumalacárregui and J.~García-Bellido,
  Phys.\ Rev.\ D {\bf 89}, 064046 (2014)

\bibitem{Langlois:2015cwa} 
D.~Langlois and K.~Noui,
JCAP {\bf 1602}, no. 02, 034 (2016)

\bibitem{Langlois:2015skt} 
D.~Langlois and K.~Noui,
JCAP {\bf 1607}, no. 07, 016 (2016)

\bibitem{Crisostomi:2016tcp} 
M.~Crisostomi, M.~Hull, K.~Koyama and G.~Tasinato,
JCAP {\bf 1603}, no. 03, 038 (2016)

\bibitem{Crisostomi:2016czh} 
M.~Crisostomi, K.~Koyama and G.~Tasinato,
JCAP {\bf 1604}, no. 04, 044 (2016)

\bibitem{Achour:2016rkg} 
J.~Ben Achour, D.~Langlois and K.~Noui,
Phys.\ Rev.\ D {\bf 93}, no. 12, 124005 (2016)

\bibitem{Ferrara:2013rsa} 
  S.~Ferrara, R.~Kallosh, A.~Linde and M.~Porrati,
  Phys.\ Rev.\ D {\bf 88}, no. 8, 085038 (2013)

\bibitem{Ade:2013zuv} 
  P.~A.~R.~Ade {\it et al.} [Planck Collaboration],
  Astron.\ Astrophys.\  {\bf 571}, A16 (2014)

\bibitem{Langlois:2017}
  D.~Langlois, R.~Saito, D.~Yamauchi and K.~Noui,
  Phys.\ Rev.\ D {\bf 97}, no. 6, 061501 (2018)

\bibitem{Crisostomi:2017}
  M.~Crisostomi and K.~Koyama,
  Phys.\ Rev.\ D {\bf 97}, no. 8, 084004 (2018)

\bibitem{Tsujikawa:2004my}
S.~Tsujikawa and B.~Gumjudpai,
Phys.\ Rev.\ D {\bf 69}, no. 12, 123523 (2004)

\bibitem{Karam:2017zno}
A.~Karam, T.~Pappas and K.~Tamvakis,
Phys.\ Rev.\ D {\bf 96}, no. 6,064036 (2017)

\bibitem{Kallosh:2013yoa} 
  R.~Kallosh, A.~Linde and D.~Roest,
  JHEP {\bf 1311}, 198 (2013)

\bibitem{Ferrara:2013rsa} 
  S.~Ferrara, R.~Kallosh, A.~Linde and M.~Porrati,
  Phys.\ Rev.\ D {\bf 88}, no. 8, 085038 (2013)

\bibitem{Garriga:99}
J.~Garriga and V.~F.~Mukhanov,
Phys.\ Lett.\ B {\bf 458}, 219 (1999)

\bibitem{Lorenz:08}
L.~Lorenz, J.~Martin, and C.~Ringeval,
Phys.\ Rev.\ D {\bf 78}, 083513 (2008)

\end{thebibliography}
\end{document}